\documentclass[aps,epsf,eqsecnum,feynmp,graphics,amsfonts,amssymb,latexsym,amsmath,twocolumn]{revtex4}
\usepackage[latin1]{inputenc}
\usepackage{graphics}
\usepackage{amsfonts, amssymb, latexsym, amsmath}
\input epsf.sty
\def\aut#1{#1}

\begin{document}
\title{End-to-End
Distribution Function
of Two-Dimensional Stiff Polymers\\ for all Persistence Lengths
}
\author{B.~Hamprecht$^1$, W.~Janke$^2$ and H.~Kleinert$^1$}
\affiliation{$^1$ Institut für Theoretische Physik, Freie Universität Berlin,\\
Arnimallee 14, D-14195 Berlin, Germany\\
$^2$ Institut für Theoretische Physik, Universität Leipzig,\\
Augustusplatz 10/11, D-04109 Leipzig, Germany\\
{\scriptsize e-mail: bodo.hamprecht@physik.fu-berlin.de}
{\scriptsize e-mail: wolfhard.janke@itp.uni-leipzig.de}
{\scriptsize e-mail: hagen.kleinert@physik.fu-berlin.de}}
\begin{abstract}
We set up and solve a recursion relation for all even moments
of a two-dimensional stiff polymer
(Porod-Kratky wormlike chain)
and determine from these moments
a simple analytic expression for
the end-to-end distribution
at all persistence lengths.
\end{abstract}
\maketitle
\section{Introduction}
In a recent
note \cite{BodoHagen},
two of us
set up a recursion relation
for
the even moments of the end-to-end distribution of stiff polymers
in $D$ dimensions and used the resulting moments
of high order
to
 construct an simple analytic
 distribution function
of the end-to-end distance ${\bf R}$.
For large persistence length $\xi$,
the result agrees
well
 with
perturbative and Monte Carlo results
of Wilhelm and Frey \cite{Frey}, for small
$\xi$ with the Daniels correction \cite{Dan}.

Recently, Dahr and Chaudhuri \cite{Dhar} have pointed out
the existence of
an interesting dip-structure in the distribution function
for two dimensions
appearing
at intermediate $\xi$-values
 in the {\em spatial\/} density function $P_L({\bf R})\equiv P_L(R)/R$
normalized to unity as $\int d^2x P_L({\bf R})=1$.
The often-plotted {\em radial\/} density function $P_L(R)$
with the normalization $\int P_L(R)~dR=1$
makes this dip almost invisible.
If one wants this dip tp
show up in a simple analytic approximation
of the type in Ref.~\cite{BodoHagen},
several more parameters are needed.
The purpose of this note
is to construct
such an analytic expression
which fits excellently
high-precision Monte Carlo data.

The end-to-end
distribution function
of a stiff polymer  in two dimensions
is
given by the path integral
 \cite{Hagen}
\begin{align}
P_L({\bf R})\! \propto  \!\!
\int d\phi_b
\, d\phi_a
 {\cal D}\phi(s) e^{ -{E_{\rm b}}/{k_B T}}\! \delta ^{(2)}\!
        \left( \!{\bf R}\!-\!\!\int_0^L\!\! ds
         {\bf u}(s)\!\right)
\label{15.82nc}
\end{align}
with
the bending energy
\begin{align}
E_{\rm b} =\frac{\kappa}2~\int _0^Lds\,[{\bf u}'(s)]^2\,.
\end{align}
where ${\bf u}(s)=(\cos\phi(s),\sin \phi(s))$
are the direction vectors of the polymer links,
and
 $\kappa $ is the stiffness
which defines
the persistence length
$\xi\equiv 2{ \kappa }/{k_BT}$.
Due to the presence of the $ \delta^{(2)}$-function
in the integrand, the path integral is not exactly
solvable.
It is, however, easy to find
arbitrarily high even moments
of the radial distribution
of the end-to-end distance
$P_L(R)$:
\begin{eqnarray}
 \left\langle { R}^{2n}\right\rangle
 \equiv
 \int dR\,{ R}^{2n}
P_L(R).
\label{@mome}\end{eqnarray}
This was explained
for  $D$-dimensions
in Ref.~\cite{BodoHagen}.
In two dimensions,
the
moments (\ref{@mome})
can be obtained
from the coefficient of $\lambda ^{2n} /2 ^{2n} (2n)!$
 in the expansion of
the integral
\begin{eqnarray}
f(\tau ;\lambda )\equiv \int_{0}^\pi d\theta \,
\psi(\theta,\tau ; \lambda ),
\label{@}\end{eqnarray}
in powers of $ \lambda $, evaluated at the euclidean time
$\tau = \xi$.
Here we use natural units with $L=1$.
  The wave function $\psi(\theta,\tau ; \lambda )$ is a solution of the
Schrödinger equation in euclidean time
\begin{eqnarray}
\hat H\psi(\theta,\tau; \lambda )=
-\frac{d}{d\tau}\psi(\theta,\tau; \lambda ),
\label{15.82nc4}
\end{eqnarray}
where
\begin{eqnarray} \!\!\!\!\!\!\!
\hat H =
 -\frac{1}{2}\Delta+\frac{1}{2}\lambda \,z
\equiv -\frac{1}{2}\frac{d^2}{d\theta^2}+
\frac{1}{2}\lambda \,\cos\theta.
\end{eqnarray}
\section{Recursive Solution of the Schroedinger Equation.}
The function
$f( L ; \lambda )$
has a spectral representation
\begin{align}
f( L ; \lambda ) \!
\equiv \!\sum_{l=0}^\infty\frac{\int_{0}^\pi d\theta\,\varphi^{(l)}{}^\dagger(\theta)
 \exp{\left(-{E^{(l)}L}\right)}
~ \varphi^{(l)}(0)}{\int_{0}^\pi d\theta\,
\varphi^{(l)}
{}^\dagger(\theta)~\varphi^{(l)}(\theta)},
\label{II1}
\end{align}
where the $\varphi^{(l)}(\theta)$ are
arbitrarily
 normalized eigensolutions of the time-independent interacting Schrödinger equation
$\hat H\varphi^{(l)}(\theta)=E^{(l)}\varphi^{(l)}(\theta)$.
Applying perturbation theory to this problem, we start from
the eigenstates
$|l\rangle $
of the
unperturbed Hamiltonian $\hat H_0=-\Delta /2$,
with eigenvalues $E_0^{(l)}=l^2/2$, and
form the
eigenfunctions $\varphi^{(l)}(\theta)=\left\langle \theta|l\right\rangle $
with the explicit form
$\varphi^{(0)}(\theta)=1/\sqrt{4\pi}$
and  $\varphi^{(l)}(\theta)=\cos(l\theta)/\sqrt{\pi}$.
Note that the ground state wave function is
not normalized to unity
 on purpose,
for later convenience.
Now we set up a recursion scheme for the expansion coefficients
 $\gamma^{(l)}_{l',i}$
and
$\epsilon^{(l)}_j$ of the eigenfunctions and their energies:
\begin{eqnarray}
|\varphi^{(l)} \rangle =\sum_{l',i=0}^\infty~\gamma^{(l)}_{l',i}~\lambda ^i~\left|l'\right\rangle,
~~
E^{(l)}
=\sum_{j=0}^\infty ~\epsilon^{(l)}_j~\lambda ^j.
\label{II3}
\end{eqnarray}
The procedure is
 described in \cite{BodoHagen,HaPe}.
The properties  of the unperturbed system for $\lambda=0$
determine the initial
conditions for the recursion:
\begin{equation}
\label{GAMMA}
\gamma_{l,i}^{(l)}=\delta_{i,0}\,, \qquad \gamma_{k,0}^{(l)}=\delta_{l,k}\,,
\qquad \epsilon _0^{(j)}=j^2/2\,.
\end{equation}
To proceed, we need the the matrix elements of the
perturbing Hamiltonian $\hat H_I$ in the basis of the
unperturbed eigenstates. They are simply
 $\langle n |\hat H_I|n \pm 1 \rangle=\lambda/2$.
Inserting the
expansions (\ref{II3})
into the Schrödinger equation (\ref{15.82nc4}), projecting the result onto some base vector $\langle k|$,
 and extracting the coefficient of $\lambda^i$, we obtain the following recursion
relations:
\begin{align}
\label{EQ1_1}
\epsilon_i^{(0)} = \gamma_{1,i-1}^{(0)},\qquad
\epsilon_i^{(l)} = (\gamma_{l+1,i-1}^{(l)}+\gamma_{l-1,i-1}^{(l)})/2 ,
\end{align}
and
\begin{align}
\label{EQ1_2}
\gamma_{0,i}^{(l)} = \frac{2}{l^2}\left(\gamma_{1,i-1}^{(l)}-\sum\limits_{j=1}^{i-1}
\epsilon_j^{(l)}\gamma_{0,i-j}^{(l)}\right),
\end{align}
\begin{align}
\label{EQ1_3}
\gamma_{k,i}^{(l)} = \frac{1}{l^2-k^2} \left(\gamma_{k+1,i-1}^{(l)}+\gamma_{k-1,i-1}^{(l)}-2\sum\limits_{j=1}^{i-1}
\epsilon_j^{(l)}\gamma_{k,i-j}^{(l)}\right)\,.
\end{align}
Starting from the initial values (\ref{GAMMA}), these recursion
relations determine
successively the
 higher-order expansion coefficients in (\ref{II3}).
Inserting the resulting
expansions (\ref{II3}) into Eq.~(\ref{II1}),
only the constant parts in $\varphi^{(l)}(\theta)$ which are independent of $\theta$
will survive the integration in the numerators. Therefore $\varphi^{(l)}(\theta)$ in the numerators of (\ref{II1}) may be replaced by the constants:
\begin{align}
\varphi^{(l)}_{\rm symm}
\equiv
\int _0^{2\pi}d\theta\,
\varphi^{(0)}{}^\dagger(\theta)
\varphi^{(l)}(\theta)
=\frac{1}{2}\sum_{i=0}
^\infty\gamma^{(l)}_{0,i}\, \lambda ^i\,,
\end{align}
the factor $1/2$ reflecting the
special normalization of $\varphi^{(0)}(\theta)
$.
The denominators of (\ref{II1}) become explicitly
\begin{align}
\int\limits_{0}^\pi d\theta\,
\varphi^{(l)}
{}^\dagger(\theta)~\varphi^{(l)}(\theta)=\sum_i
\left(|\gamma_{0,i}^{(l)}|^2/2 + \sum_{l'}|\gamma_{l',i}^{(l)}|^2\right)
 \lambda^{2i},
\end{align}
where the sum over $i$ is limited by the
power of $\lambda^2$ up to which we want to carry the perturbation series;
also $l'$ is restricted to a finite number of terms,
because of a band-diagonal structure of the $\gamma_{l',i}^{(l)}$
(see \cite{BodoHagen}).
\noindent
Extracting the coefficients of the power expansion in $ \lambda $ from
(\ref{II1}) we
obtain all desired  moments
of the end-to-end distribution,
the
lowest two being:
\begin{equation}
\!\!\!\!\!\!
\!\!\!\!\!\!\!\!
\!\!\!\!\!\!\!\!
\!\!\!\!\!\!\!\!
\!\!\!\!\!\!\!\!
\langle R^2\rangle = {2} \left\{
\xi L-{\xi ^2} \left[ 1-e^{-L/\xi }\right] \right\},
\label{II5}
\end{equation}
\begin{align} \label{R4}
\langle R^4\rangle
=&\,8L^2 \xi ^2 - L\xi ^3\left(30+\frac{40}{3}e^{-L/\xi}\right)
~~~~~~~~~~\nonumber \\
+&\xi ^4
\left(
\frac{87}{2}
-\frac{392}{9}e^{-L/\xi}+\frac{1}{18}e^{-4L/\xi}
\right).
\end{align}
The calculation of higher moments
is straightforward
with
a Mathematica
program, which we have placed
on the internet
in notebook form
\cite{NBK}.
The above lowest moments agree with those in
Ref.~\cite{Dhar}.
\section{End-to-End Distribution}
\begin{figure}[htp!]
\begin{center}
\setlength{\unitlength}{1cm}
\begin{picture}(8.5,5.5)
\put(0,0){\scalebox{.8}[.8]{\includegraphics*{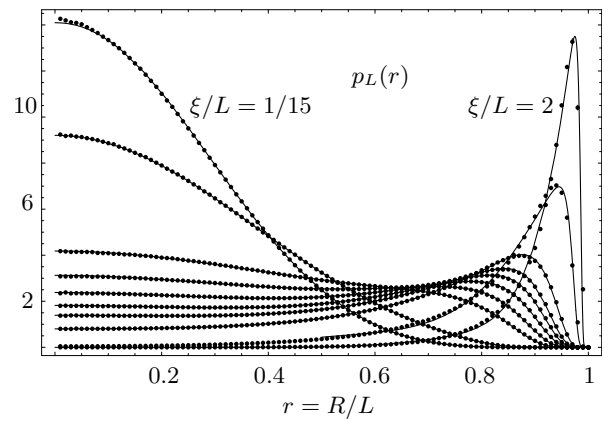}}}
\put(2.1,0.1){0.2}
\put(3.5,0.1){0.4}
\put(4.9,0.1){0.6}
\put(6.3,0.1){0.8}
\put(7.88,0.1){1}
\put(0.3,1.1){ 2}
\put(0.3,2.4){ 6}
\put(0.3,3.7){10}
\put(3.88,-0.3){$r=R/L$}
\put(4.8,4.1){$p_L(r)$}
\put(2.65,3.7){$\xi/L=1/15$}
\put(6.35,3.7){$\xi/L=2$}
\end{picture}
~\\
~\\[-1em]
\caption[D2]{End-to-end distribution $p_L(r)$ in $D=2$
dimensions as a function of $r=R/L$ for various values
of the stiffness ${\xi}/L=1/15,\,1/10,\,1/5,\,1/4,\,3/10,\,7/20,\,2/5,\,1/2,\,1,\,2$. The solid curves show the model functions (\ref{modell}) with parameters from Table~\ref{T2}. The dots
show the Monte Carlo data.
 }
\label{D2}
\end{center}
\end{figure}

\begin{figure}[htp!]
\begin{center}
\setlength{\unitlength}{1cm}
\begin{picture}(8.5,6)
\put(0,0){\scalebox{.8}[.8]{\includegraphics*{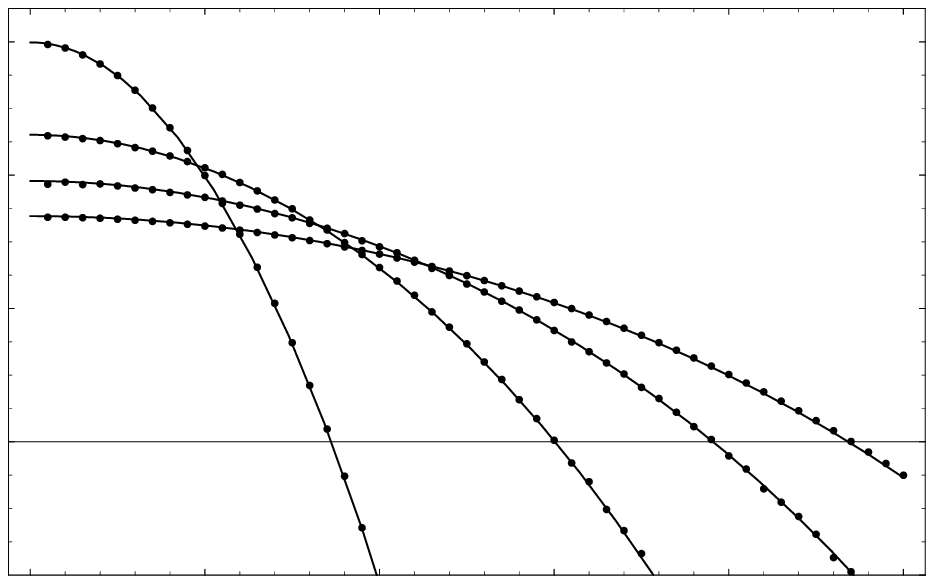}}}
\put(3.25,0.1){0.2}
\put(6.1,0.1){0.4}
\put(0.2,1.5){0}
\put(0.2,2.55){2}
\put(0.2,3.6){4}
\put(.8,2.3){$\xi/L=1/400$}
\put(6.15,2.3){$\xi/L=1/30$}
\put(4.2,-0.1){$r=R/L$}
\put(4.8,4.1){$\log p_L(r)$}
\end{picture}
\caption[D2L]{
End-to-end distribution near ${\bf R}=0$
in $D=2$ dimensions plotted logarithmically
as a function of $r=R/L$
for various values of the stiffness for
floppy polymers with
${\xi}/L=1/400,\,1/100,\,1/50,\,1/30$. The solid curves show the model functions (\ref{modell}) with parameters from Table~\ref{T2}. The dots represent Monte Carlo data.
 }
\label{D2L}
\end{center}
\end{figure}

As in the previous paper,
we shall now
set up an
analytic  distribution function
for $p_L(r)\equiv
P_L({\bf R})$, where $r=R/L$:
\begin{align} \!\!\!\!
p_L(r)=(a_0\!+\!a_2x^2\!+\!a_4x^4\!+\!a_6x^6)x^k(1-x^\beta)^m\,.
\label{modell}
\end{align}
The
parameters $a_0,\dots,a_6,\,k,\,\beta,\,m$
are functions of $\xi/L$ and are
determined by forcing the moments of
(\ref{modell}),
\begin{align}
\displaystyle\left\langle { R}^{2n}\right\rangle=L^{2n}
\int_0^1 r^{2n+1}\,p_L(r)\,dr\,,
\end{align}
 to fit the exact moments in the
range $0 \leq n \leq {\rm Max}(6,10\,{\xi}/L)$.
For floppy chains with $\xi<1$ we set $k=0$;
for very floppy chains with $\xi<1/20$ the choice $a2=a4=a6=0$
guarantees excellent accuracy.
 A comparison of $p_L(r)$ with Monte-Carlo data is shown in Fig.~\ref{D2},
and with higher resolution near ${\bf R}=0$ in Fig.~\ref{D2L}. The
associated coefficients are listed in Table~\ref{T2}.
\begin{table*}
\begin{tabular}{||r|r|r|r|r|r|r|r||}
\hline \hline
$\hat{\xi}$ & $a_0$ & $a_2$  & $a_4$  & $a_6$ & $k$  & $m$ & $\beta$ \\
\hline \hline
$1/400$ & $400.0$  &$0$ & $0$ & $0$ & $0$ & $196.784$   & $1.99496$   \\
\hline
$1/100$ & $100.0$  &$0$ & $0$ & $0$ & $0$ & $47.5378$   & $1.98197$   \\
\hline
$1/50$ & $50.0$  &$0$ & $0$ & $0$ & $0$ & $22.9930$   & $1.97564$   \\
\hline
$1/30$ & $29.5302$  &$-58.9195$ & $77.9373$ & $-87.3526$ & $0$ & $12.0224$   & $2.00505$   \\
\hline
$1/15$ & $14.0952$  &$-29.8504$ & $66.8842$ & $-68.1985$ & $0$ & $5.62896$   & $2.21169$   \\
\hline
$1/10$ & $9.20629$  & $-34.7515$ & $50.6223$ & $-26.2289$ & $0$ & $13.2737$  & $10.5486$   \\
\hline
$1/5$  & $4.18239$  &$-7.45808$ & $11.616$ & $-7.30855$ & $0$  & $10.2031$  & $16.6444$    \\
\hline
$1/4$  & $3.12655$  &$-4.9930$ & $13.1086$ & $-10.0222$ & $0$  & $9.42195$   & $20.0750$   \\
\hline
$3/10$ & $2.38054$  &$-3.38168$ & $12.8823$ & $-9.51483$ & $0$  & $9.16782$   & $23.0164$   \\
\hline
$7/20$ & $1.82132$  &$-2.062292$ & $11.2343$ & $-7.24306$ & $0$  & $8.84230$   & $25.5206$   \\
\hline
$2/5$  & $1.39171$  &$-0.952158$ & $8.94986$ & $-4.33545$ & $0$  & $8.49552$   & $27.9120$   \\
\hline
$1/2$  & $0.800939$  &$0.647524$ & $4.36711$ & $1.36933$ & $0$  & $8.06681$   & $33.2814$   \\
\hline
$1$    & $42.8376$  &$-173.308$ & $263.327$ & $-123.515$ & $4.9880$ & $11.4933$   & $85.8428$  \\
\hline
$2$   & $504.624$  &$-1832.52$ & $2271.51$ & $-925.829$ & $13.4792$  & $30.4949$  & $244.143$  \\
\hline
\hline
\end{tabular}
\caption{Coefficients of the analytic
distribution
function for $p_L(r)$ in Eq.~(\ref{modell}) for
various  values of the persistence length
$\hat{\xi}$. They
are obtained by making six or seven even  moments of  $p_L(r)$
 agree with the exact ones. }
\label{T2}
\end{table*}
The calculation of
the coefficients in (\ref{modell})
requires some care
 to
guarantee sensitivity to
possible local minima,
and to avoid
running  into unphysical oscillations.
The latter may arise from the
 existence of polynomials
in which
all moments lower than some $n$ vanish.
Such oscillations are avoided by controlling
the high moments and using only
low polynomial
coefficients in (\ref{modell}).
 For $\xi>1/20$, a more involved strategy is necessary to avoid
low-quality local solutions. We proceed as follows:
\begin{itemize}
\item
In a first step we set $a_2=a_4=a_6=0$, $\beta=2$, and determine
preliminary values for $k$ and $m$ by fitting
 two higher moments of $n$ near $10\,{\xi}/L$. The first
coefficient $a_0$ is  fixed by normalization.
This gives a reasonable starting value for $m$.
\item
In a second step,
we introduce one more of the higher moments and improve
the solution for $k$, $m$, and $\beta$.
\item
Next we solve for the coefficients $a_j$ by bringing yet more moments into play.
If ${\xi}/L < 1$, we take $k=0$ and solve for the $a_j$, keeping $\beta$ and $m$ fixed, based on four properly chosen moments.
Then we solve for  $\beta$ and $m$ keeping  the $a_j$ fixed,
based on a choice of two moments.
This alternating procedure is repeated three times.
Finally, we solve for the $a_j$, $\beta$, and $m$ simultaneously,
based on six properly chosen moments.
\item
For ${\xi}/L \ge 1$, we proceed similarly, but allow for
 $k\ne 0$. The search for good
coefficients
 $a_j$ alternating with a search for good $k$, $\beta$,
 and $m$ is repeated until it converges, with no further attempt to
solve once more for
all seven parameters simultaneously.
\end{itemize}

~\\
\begin{table}[h]
\begin{tabular}{||c|c|c|c||}
\hline \hline
${\xi}/L$& $\Sigma$ & $\Delta_{\rm abs}$ & $\Delta_{\rm rel}(N_{\rm max})$ \\
\hline \hline
$1/400$ & $3\times 10^{-12}$ & $0.000\,126$   & $0.9\%(16)$ \\
\hline
$1/100$ & $2\times 10^{-13}$ & $0.000\,033$   & $8\%(16)$  \\
\hline
$1/50$ & $1\times 10^{-10}$ & $0.000\,071$   & $1\%(16)$  \\
\hline
$1/30$ & $5\times 10^{-9}$ & $0.000\,717$   & $8\%(16)$ \\
\hline
$1/15$ & $2\times 10^{-6}$ & $0.000\,057$   & $0.9\%(16)$                         \\
\hline
$1/10$ & $5\times 10^{-5}$  & $0.000\,038$   & $0.8\%(24)$                         \\
\hline
$1/5$  & $4\times 10^{-5}$  & $0.000\,048$    & $0.5\%(24)$                         \\
\hline
$1/4$  & $9\times 10^{-5}$  & $0.000\,048$    & $0.4\%(24)$                         \\
\hline
$3/10$ & $13\times 10^{-5}$ & $0.000\,047$   & $0.3\%(24)$                          \\
\hline
$7/20$ & $2\times 10^{-4}$  & $0.000\,087$   & $0.8\%(36)$                          \\
\hline
$2/5$  & $2\times 10^{-4}$  & $0.000\,100$    & $0.8\%(36)$                         \\
\hline
$1/2$  & $2\times 10^{-4}$  & $0.000\,139$    & $0.8\%(36)$                         \\
\hline
$1$    & $2\times 10^{-4}$  & $0.000\,217$      & $0.4\%(48)$                       \\
\hline
$2$    & $8\times 10^{-5}$  & $0.004\,705$      & $1.6\%(48)$                       \\
\hline
\hline
\end{tabular}
\caption{To illustrate
the accuracy of our
analytic approximation
(\ref{modell}) of the end-to-end
distribution
 we list the quantity $\Sigma$ of Eq.~(\ref{sigma})
which measures the
deviation
of the even moments
 from the exact ones.
The other two columns
illustrate the accuracy of the Monte Carlo data
for the end-to-end distribution
by listing
the maximal deviation $\Delta_{\rm abs}$ of its moments
and the relative
 deviation $\Delta_{\rm rel}(N_{\rm max})$ up to the moment $N_{\rm max}$.
}
\label{T1}
\end{table}

We check
the quality of
our simple distribution function (\ref{modell})
with the parameters
of Table~\ref{T2}
by calculating its moments and comparing them with the exact ones.
The comparison is shown
in Table~\ref{T1}
for a large range of the persistence length $\xi$.
As a measure of the quality of approximation
we use
the quantity
 $\Sigma$,
listed in the second column of Table~\ref{T1},
 which sums up all squared deviations of the
moments of the model from the
exact ones in a relevant range of $\xi$:
\begin{align}
\label{sigma}
\Sigma(\xi)=\sqrt{\sum_{n=0}^N \left(
 \langle R^{2n}\rangle_{p_L}-
\langle R^{2n}\rangle_{P_L} \rangle
\right)^2}\,,
\end{align}
where we have extended
the sum up to the  moments
of order $N=12$ for ${\xi}/L<1/5$, and  $N=24$ for ${\xi}/L \geq 1/5$.

Let us also convince ourselves
of the high accuracy
of our Monte Carlo data
for the end-to-end distribution
in Figs.~\ref{D2}
and \ref{D2L}
by listing
the maximal deviation
\begin{align}
\Delta_{\rm abs}=\sup_{n=0}^\infty \left|
 \langle R^{2n}\rangle_{MC}-
\langle R^{2n}\rangle_{P_L} \rangle
\right|
\end{align}
of its moments
and the relative
 deviation
\begin{align}
\Delta_{\rm rel}(N_{\rm max})=\sup_{n=0}^{N_{\rm max}} \left|
 \langle R^{2n}\rangle_{MC}/
\langle R^{2n}\rangle_{P_L} \rangle-1
\right|
\end{align}
 up to the moment $N_{\rm max}$.

Remarkably,
in spite of the simplicity
of the model,
it is a nontrivial task
to obtain accurate Monte Carlo results
for $P_L({\bf R})$ near ${\bf R}=0$ which
are sensitively displayed in the plots of Fig.~\ref{D2}
but which are almost ignored by the moments.
The reason for the difficulty is the small configuration space for
the small-${\bf R}$
data since  the binning of the data is done in $R$
to estimate the density $P_L({\bf R})$.
One is caught in the
competition between large systematic errors resulting from
a too large bin size
$\Delta R$, and statistical errors
from a too small $\Delta R$.
 As a compromise
we employed in our simulations a uniform
bin size $\Delta R/L = 0.01$ which in combination with
a single-cluster update procedure and a statistics
of $10^8$ sampled chains yielded satisfactory accuracy
near
${\bf R}=0$, as shown in the high-resolution plot in Fig.~\ref{D2L}.

\vspace{.7cm}
~\\Acknowledgment\\
This work was partially supported by
ESF COSLAB Program and by the Deutsche Forschungsgemeinschaft
under Grant Kl-256. 
WJ acknowledges partial support by the German-Israel-Foundation (GIF) under
contract No.~I-653-181.14/1999.
%
%
%

%

%

\end{document}